\documentclass[preprint,12pt]{elsarticle}

\usepackage{amssymb}
\usepackage{amsmath,bm}

\usepackage{float}

\usepackage{caption}
\usepackage{subcaption}
\usepackage{graphicx}

\usepackage{tikz}
\usepackage{url}
\usepackage[margin=1in]{geometry}
\usetikzlibrary{positioning, arrows.meta}

\journal{arXiv}

\begin{document}

\begin{frontmatter}



\title{Physics-guided denoiser network for enhanced additive manufacturing data quality}

\author[inst1]{Pallock Halder}

\author[inst1]{Satyajit Mojumder\corref{cor1}}
\cortext[cor1]{Corresponding author}
\ead{satyajit.mojumder@wsu.edu}

\affiliation[inst1]{organization={School of Mechanical and Materials Engineering, Washington State University},
            addressline={}, 
            city={Pullman},
            postcode={99164}, 
            state={WA},
            country={USA}}

\begin{abstract}
Modern engineering systems are increasingly equipped with sensors for real-time monitoring and decision-making. However, the data collected by these sensors is often noisy and difficult to interpret, limiting its utility for control and diagnostics. In this work, we propose a physics-informed denoising framework that integrates energy-based model and Fisher score regularization to jointly reduce data noise and enforce physical consistency with a physics-based model. The approach is first validated on benchmark problems, including the simple harmonic oscillator, Burgers’ equation, and Laplace’s equation, across varying noise levels. We then apply the denoising framework to real thermal emission data from laser powder bed fusion (LPBF) additive manufacturing experiments, using a trained Physics-Informed Neural Network (PINN) surrogate model of the LPBF process to guide denoising. Results show that the proposed method outperforms baseline neural network denoisers, effectively reducing noise under a range of LPBF processing conditions. This physics-guided denoising strategy enables robust, real-time interpretation of low-cost sensor data, facilitating predictive control and improved defect mitigation in additive manufacturing.
\end{abstract}




\end{frontmatter}


\section{Introduction}
Advanced manufacturing systems increasingly depend on real-time sensing to monitor fabrication processes, detect anomalies, identify defects, and enable in-situ control strategies for producing high-quality, defect-free parts~\cite{sajadi2024realtime2dtemperaturefield, bhandarkar2025real, RICHTER2021100060}. In laser powder bed fusion (LPBF), for example, high-fidelity sensors like synchrotron-based X-ray imaging can capture detailed insights into melt pool behavior, microstructure evolution, and defect formation~\cite{ren2023machine, rensubmilli2024, mutswatiwa2024ultrasonic}. However, such systems are prohibitively expensive, generate vast amounts of data, and require complex infrastructure, making them unsuitable for routine industrial or small-scale applications. On the other hand, low-cost sensors—such as acoustic emission sensors~\cite{akhavan2024realtime, lialuminum2023}, photodiodes~\cite{cartermeltpool2023, JEONG2022100101}, and thermal cameras~\cite{Chen2022_ThermalImaging}—offer a more scalable and accessible alternative, but they produce inherently noisy~\cite{Dominik2022_CoaxialLPBF}, low-resolution data with limited interpretability~\cite{Dominik2024_InterlayerLPBF}. This trade-off between sensor affordability and data quality creates a significant challenge for widespread adoption of real-time monitoring in AM.

Noisy data is inherent in any experimental measurement system due to limitations in sensor fidelity, data acquisition frequency, and inherent uncertainty of the system. Various statistical techniques have been employed for noise reduction, including smoothing methods~\cite{mahmoodinoisereduction2006, SHTAYAT2024110} like moving averages, exponential smoothing, and locally weighted regression; outlier removal techniques such as Z-score analysis~\cite{Aggarwal2019_ZScore}, interquartile range filtering, and isolation forests; filtering methods like Gaussian filters, median filtering, and Fourier transform filtering, wavelet filtering~\cite{thompson2014_LLNL}; and dimensionality reduction techniques such as Principal Component Analysis (PCA)~\cite{SPIEGELBERG2017_PCA} and Singular Value Decomposition (SVD)~\cite{SAMANN2024_IFAC, Ji2022_SVD}. Additionally, statistical modeling approaches like Kalman filters~\cite{kalman1960new, Park2019_Kalman} and Bayesian methods~\cite{camerlingo2023bayesian} have also been used. However, these methods rely heavily on the observed data and lack physics guidance, which can be detrimental in real-time scenarios—such as predicting the next-layer printing parameters in additive manufacturing (AM). Machine learning offers a promising alternative to statistical approach for denoising and enable near real-time predictive capability. Techniques like autoencoders~\cite{DENG2024108685}, convolutional neural networks (CNNs)~\cite{Gondara2016_Autoencoders}, generative adversarial networks (GANs)~\cite{bengio2013generalized}, recurrent neural networks (RNNs)~\cite{ZHAO2021227}, and diffusion models~\cite{Zhu2023_DAS-VSP} have been used for data denoising. The seminal works by Lagaris et al.~\cite{Lagaris1998_PINN} and Raissi et al.~\cite{RAISSI2019686} introduced Physics-Informed Neural Networks (PINNs), which incorporate governing equations and boundary conditions into neural networks to guide learning with physical laws. Despite their promise, PINNs face challenges in optimizing complex loss landscapes and handling high-dimensional data, particularly in the presence of noise~\cite{Park2023_C-HiDeNN}. Techniques such as regularization and hybrid approaches have improved their robustness and expanded their applications from solid mechanics~\cite{hupinn2024, haghighatpinn2021} to fluid dynamics~\cite{cai2021pinnsreview} and advanced manufacturing ~\cite{sajadi2024realtime2dtemperaturefield, kats2022physicsinformed, zamiela2024physicsinformed} and weather prediction~\cite{moreno2024pinnweather}. However, noise and difficulty in calibrating the governing PDEs parameters remain significant hurdles. To address this, we propose a physics-informed denoising framework that integrates Energy-Based Models (EBMs)~\cite{lecun2006energybased} with Fisher score-based~\cite{osborne1992fishers} regularization to formulate a loss function that accounts for both noisy data and physical constraints. Previous work of Pilar and Wahlstrom ~\cite{pmlr-v242-pilar24a} has combined PINNs with EBMs for denoising in fluid dynamics, such as the Navier–Stokes equation, but to our knowledge, Fisher score techniques have not been used in this context. EBMs focus on probability density estimation which is not a direct indicator of feature importance~\cite{gustafsson2020energy}.  Fisher Score is a commonly used supervised feature selection method that directly ranks features according to their ability to discriminate between classes. It calculates a ratio of between-class variance to within-class variance for each feature, allowing for a clear ranking based on the statistical difference between class means. While EBMs have proven effective, estimating their energy function can be computationally challenging, especially in complex models like neural networks. Conversely, the Fisher Score, being a simpler statistical measure, is generally easier and faster to compute~\cite{Gu2011_FisherScore, Liu2024_FisherScore}. 
Thermal sensor data in LPBF, particularly from the melt pool, is notoriously noisy due to indirect measurement limitations and the extreme thermal conditions. Nevertheless, this thermal data is crucial as it governs melt pool dynamics, microstructure evolution, and defect formation. While high-fidelity simulations using computational fluid dynamics (CFD)~\cite{li2024statistical, sjitlpbf2023, ganbm2019} or finite element analysis (FEA)~\cite{leonor2024gomelt} can reasonably predict temperature fields, they are computationally expensive and impractical for real-time applications. Surrogate models~\cite{sarker2025data, guo2025tensordecompositionbasedpriorisurrogatetaps} and reduced-order approaches~\cite{lu2023convolution, lu2023extendedtensordecompositionmodel} have been proposed to accelerate these simulations, but they require extensive calibration with experimental data. Recently, Tang et al.~\cite{TANG2023116197, TANG2024104574} demonstrated a PINN framework for simulating single-track LPBF, showing fast inference times after training and verified against a finite element-based solver. However, the results are not validated against a real experiment as experimental data are noisy and a deterministic simulation are not able to capture such noise. In this context, Li et al. \cite{li2024statistical} used extensive calibration of their statistical model to develop a digital shadow of LPBF and showed that a statistical calibration can be helpful for the model to simulate the LPBF process without directly accounting the experimental noise. However, the statistical calibration process is expensive and not suitable for real-time prediction and control of the process. For such digital twin models, noisy data can significantly degrade model robustness and prediction accuracy, ultimately impacting print quality.

To address this challenge of experimental noisy data, we developed a physics-informed denoising network that combines statistical noise reduction via EBM or Fisher score-based regularization. The architecture was first validated on the benchmark problems including simple harmonic oscillator, Burgers' equation, and Laplace’s equation. We then applied it to LPBF by developing a PINN model for the AM process and using it to guide the denoising of experimental thermal emission planck data (TEP) which is an indicator of average melt pool temperature collected using a photodiode-based sensor. Our results show that the proposed physics-guided denoiser performs robustly across a wide range of LPBF conditions and outperforms a standard neural network denoiser. This enables real-time noise reduction of AM data and supports predictive control for defect mitigation and quality improvement in printed parts. The paper is organized as follows: Section 2 provides a physics-guided denoiser networks and its individual blocks. Section 3 uses the denoiser networks for three numerical benchmark problems. Section 4 describes the PINN model for LPBF simulation which is used to denoise real experimental measurement. Section 5 discusses the experimental data collection process, and Section 6 presents the performance of the denoiser network on synthetic and real experimental noisy data for LPBF. Finally, a conclusion is presented with future works in Section 7. 

\section{Physics-informed machine learning for noise reduction}
\begin{figure}[ht]
    \centering
    \includegraphics[width=1.0\linewidth]{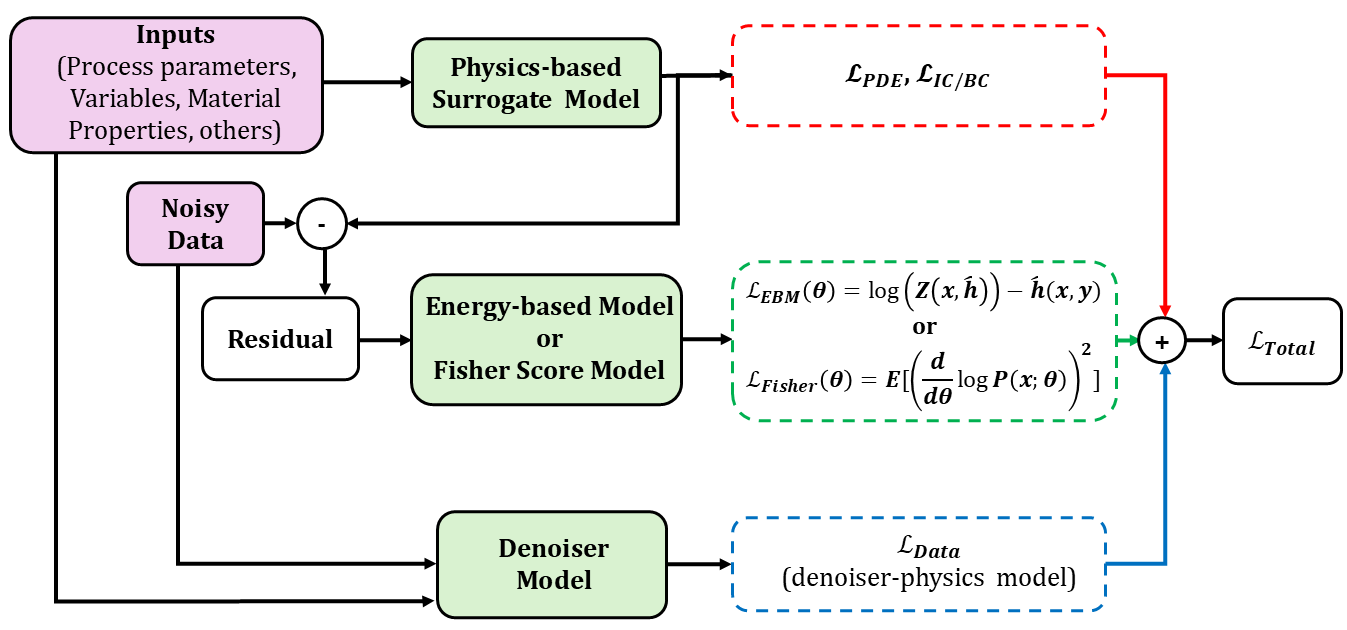}
    \caption{Overview of the proposed physics-informed denoiser framework combining a PINN, data consistency loss, and statistical modules (EBM and Fisher Score). All components contribute to the total loss \(\mathcal{L}_{\text{TotalLoss}}\) used for joint training.}
    \label{fig:fig1_framework}
\end{figure}

The proposed denoiser framework combines physics-based supervision with data-driven statistical modeling to enhance prediction accuracy under noisy measurement conditions. It incorporates a PINN, which enforces the model's consistency with physical laws by minimizing loss terms based on the governing partial differential equations (PDE) and initial/boundary conditions (ICs/BCs). To account for noise, residuals or differences between the noisy data and the physics-based predictions are passed into one of two probabilistic regularizers: the Energy-Based Model (EBM), which imposes statistical structure through a log-partition-based loss (\(\mathcal{L}_{\text{EBM}}\)); or the Fisher Score Model, which measures the sensitivity of governing variables by a Fisher Information-based loss (\(\mathcal{L}_{\text{Fisher}}\)). In parallel, a denoiser model receives both physical input features similar to a PINN model and experimentally noisy data to produce denoised predictions. A data loss (\(\mathcal{L}_{\text{Data}}\)) is computed between the denoiser model's output and the PINN prediction to guide learning process. All individual loss terms are combined into a total loss function (\(\mathcal{L}_{\text{Total}}\)), which is used to jointly train the hybrid framework. This integrated approach allows robust and physical consistency for denoising across varying noisy conditions, , as illustrated in Figure~\ref{fig:fig1_framework}. The details of the individual model in the framework is described in the following section. 

\subsection{Physics-informed neural network (PINN)}
According to universal approximation theory, a deep neural network (DNN) can approximate any continuous function to an arbitrary degree of accuracy with a sufficient number of neurons and training data. DNNs are data hungry and require large amounts of labeled data for training. PINNs offer an advantage to the DNNs by embedding prior knowledge of governing physics, which reduces the dependency on large datasets and constraining the prediction on a feasible solution region. PINNs minimize the total loss by combining data-driven errors with residuals derived from governing differential equations. Unlike conventional numerical solvers, PINNs are mesh-free approach, making them suitable for high-dimensional and complex systems where traditional solvers struggle due to computational costs or effects of increasing dimensionality. 
PINN is an effective way to integrate the physics of the problem in the denoiser network. While PINN uses the governing equations, and boundary and initial conditions to guide the solution, it can help denoising by identifying the physical trend from the noisy data. Here, we use PINN as a physics-based surrogate model which can be replaced by other physics-based simulation models as well. PINN is a natural choice for our framework to build a unified denoiser network. 

\subsection{Denoiser model}
The denoiser model is implemented as a fully connected feedforward neural network (FFNN), designed to recover clean signals from noisy observations by learning mappings informed by both physical and contextual inputs. It takes as input a combination of spatiotemporal coordinates, relevant problem-specific parameters, and the corresponding noisy measurements. The model outputs a denoised estimate of the underlying true quantity of interest. This generic formulation allows the denoiser to be easily adapted to different physical systems and types of noise. Depending on the problem, the architecture, number of layers, and input features can be customized accordingly. We used specific architectural choices, training, and evaluation strategies for each problem in the following sections.

\subsection {Energy-based model (EBM)}
EBM is a powerful method to learn probability densities  from the data. It provides the flexibility of a neural network and can be used for the regression problem. EBM learns the complex relationships between input variables \(x\) and output variables \(y\) by defining an energy function \(F(x, y)\). Generally, traditional neural networks maps the inputs directly to the outputs. But, EBMs associate each possible pair of \((x, y)\) with a scalar energy value. The fundamental principle of EBM is that lower energy values correspond to more likely or compatible pairs, while higher energy values represent less probable or incompatible pairs. The objective of EBMs is to learn an energy function \(F(x, y)\) that correctly assigns low energy to high likely outputs and high energy to low likely outputs. For the proposed framework, the EBM serves as a statistical regularizer that penalizes implausible denoised outputs based on a learned energy landscape, enabling the denoiser to recover physically meaningful data from noisy observations. 
In this study, the EBM is using the following parameterization:

\begin{equation}
    P(x, y; \theta ) = \frac{\exp{(F(x, y))}}{Z}\
\end{equation}

Here,  \( P(x, y; \theta) \) is the probability distribution over the input \( x \) and output \( y \), and learned using a neural network parameterized by \( \theta \). \( F(x, y) \) is the energy function, which assigns a scalar energy value to each pair \( (x, y) \). Lower energy corresponds to more likely pairs, while higher energy corresponds to less likely ones. The exponent ensures that lower-energy configurations are assigned higher probability. To ensure  that \( P(x, y; \theta) \) sums (or integrates) to 1, partition function or normalization constant \( Z \) is used. To calculate the partition function, we can use the continuous case, where \( Z \) is defined as: 

\begin{equation}
    Z = \int_{x, y} \exp(F(x, y)) \, dx \, dy
\end{equation}

However, computing this integral exactly is often intractable in practice. Therefore, we approximate the integral using a discrete sum over sampled pairs \( (x, y) \):

\begin{equation}
    Z \approx \sum_{x, y} \exp(F(x, y))
\end{equation}

For optimizing the loss function for this combined model, the Negative Log-Likelihood (NLL) or log-partition-based loss (\(\mathcal{L}_{\text{EBM}}\)) was introduced, which is described as follows:

\begin{equation}
    NLL = -\log(P(x, y; \theta)) = log (Z) - F(x,y)
\end{equation}

\subsection{Fisher-score model}
The Fisher score and the Fisher information matrix are fundamental concepts in statistics and machine learning, widely used in maximum likelihood estimation (MLE)~\cite{jaketae_fisher}. These tools quantify the sensitivity of the likelihood function to changes in the model parameters and reflect the amount of information a parameter carries about the observed data. The Fisher score, also known as the score function, represents the gradient of the log-likelihood function with respect to a parameter \( \theta \). For a given probability density function \( P(x; \theta) \), the log-likelihood for a dataset \( X = \{ x_1, x_2, \dots, x_n \} \) is given by:
\begin{equation}
    \mathcal{L}(\theta) = \sum_{i=1}^{n} \log P(x_i; \theta)
\end{equation}
The corresponding Fisher score function is defined as:
\begin{equation}
    S(\theta) = \frac{d}{d\theta} \sum_{i=1}^{n} \log P(x_i; \theta)
\end{equation}
Solving \( S(\theta) = 0 \) provides the critical points of the log-likelihood function and is a common approach to obtain the MLE of the parameter \( \theta \).

The Fisher information quantifies the amount of information that an observable variable \( x \) carries about an unknown parameter \( \theta \). It is defined as the expected value of the squared Fisher score:
\begin{equation}
    I(\theta) = \mathbb{E} \left[ \left( S(\theta) \right)^2 \right]
\end{equation}
This formulation allows the Fisher score method to play a central role in parameter optimization and score the data based on their noise level. In the proposed denoiser framework, the Fisher score loss encourages the model to learn denoising patterns that are not only accurate but also statistically informative for denoising purposes.

\section{Numerical examples}
To test the physics-guided denoiser network, three sets of numerical experiments are conducted using well-established benchmark problems: i) Simple Harmonic Oscillator (SHO), ii) 1D time-dependent Burgers’ equation, and iii) 2D Laplace equation. These equations are frequently used to assess the accuracy, generalization, and robustness as well as the performance of the proposed numerical method as the exact or reference solutions of the governing equations are known. We first established a PINN model for these equations and validated it against the analytical solution. Further, we introduced synthetic noise with the analytical solution data for the problems to examine the denoising capabilities of the proposed denoiser network. This enables a scope for comparing and evaluating the models and their performances under varying noisy conditions where the ground truth is known.

\subsection{Simple Harmonic Oscillator}
The Simple Harmonic Oscillator (SHO) is chosen as a baseline problem because its analytical solution is well-known, allowing direct evaluation of the denoiser model’s accuracy. The classical SHO system describes the motion of a particle of mass \(m \) attached to a spring with spring constant \(k \), and is governed by:

\begin{equation}
m \frac{d^2 x}{dt^2} + kx = 0,
\end{equation}

where \(x(t) \) denotes the displacement of the oscillator at time \(t\). This second-order linear homogeneous differential equation models periodic motion, making it ideal for assessing baseline model performance. The damping force is not considered here. 

For simplicity, and without loss of generality, the normalized case was considered where \( \frac{k}{m} = 1 \), which simplifies the equation to:
\begin{equation}
    \frac{d^2 x}{dt^2} + x = 0.
\end{equation}
The initial conditions for this system were assumed as
\begin{equation}
    x(0) = 1, \quad \frac{dx}{dt}(0) = 0.
\end{equation}
The general solution of the differential equation can be written as the linear combination of sine and cosine functions:
\begin{equation}
    x(t) = A \cos(t) + B \sin(t),
\end{equation}
where \( A \) and \( B \) are constants determined by the initial conditions.

\begin{figure}
    \centering
    \includegraphics[width=0.9\linewidth]{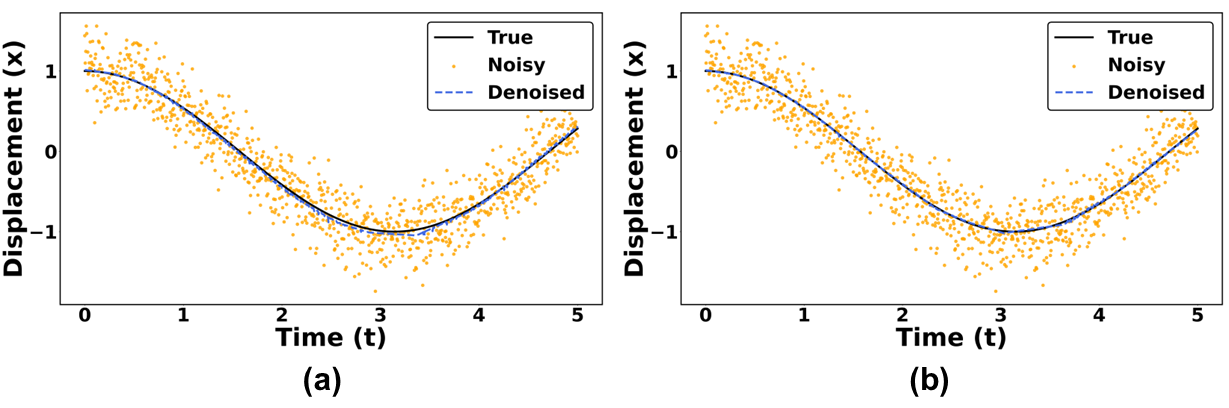}
    \caption{Comparison of SHO predictions from (a) Denoiser-EBM and (b) Denoiser-Fisher models.}
    \label{fig:sho-denoisers}
\end{figure}

In the real world, measurements of such displacement (using a camera) are often subject to noise, sensor inaccuracies, or external disturbances. To replicate these conditions, up to 25\% random Gaussian noise was introduced to the exact solution, creating a synthetic noisy dataset. These datasets were then used for the training of the physics-informed denoiser network.  The solution of the denoiser network is compared to a Vanilla neural network(a simple feedforward neural network trained using MSE loss for regression)  and a PINN. To evaluate the effectiveness of the proposed denoising strategies, the predictions obtained from the Denoiser-EBM and Denoiser-Fisher models are compared side by side in Figure~\ref{fig:sho-denoisers}. Both models aim to recover the true solution from noisy observations while adhering to the underlying physical laws.

Evaluating the comparative performance of the Vanilla, PINN and denoiser-models is essential to understand their robustness under varying noise conditions. RMSE and \(R^2 score\) metrics are used to compare these models (see Table \ref{tab:model_noise_comparison}). To simulate different noise levels, Gaussian noise with zero mean and varying standard deviations (5\% to 25\% of the true temperature values) was added to the data.
\begin{table}[htbp]
\centering
\renewcommand{\arraystretch}{1.3}
\resizebox{\textwidth}{!}{%
\begin{tabular}{|c|cc|cc|cc|cc|}
\hline
\textbf{Noise Level} & \multicolumn{2}{c|}{\textbf{Vanilla}} & \multicolumn{2}{c|}{\textbf{PINN}} & \multicolumn{2}{c|}{\textbf{Denoiser-EBM}} & \multicolumn{2}{c|}{\textbf{Denoiser-Fisher}} \\
\cline{2-9}
 & \textbf{RMSE} & $\mathbf{R^2}$ \textbf{score} & \textbf{RMSE} & $\mathbf{R^2}$ \textbf{score} & \textbf{RMSE} & $\mathbf{R^2}$ \textbf{score} & \textbf{RMSE} & $\mathbf{R^2}$ \textbf{score} \\
\hline
5\%  & 0.05183 & 0.99384 & 0.00474 & 0.99995 & 0.02667 & 0.99837 & 0.00843 & 0.99984 \\
10\% & 0.10160 & 0.97633 & 0.00474 & 0.99995 & 0.02678 & 0.99836 & 0.00853 & 0.99983 \\
15\% & 0.15154 & 0.94736 & 0.00474 & 0.99995 & 0.02697 & 0.99833 & 0.00867 & 0.99983 \\
20\% & 0.20111 & 0.90728 & 0.00474 & 0.99995 & 0.02741 & 0.99828 & 0.00887 & 0.99982 \\
25\% & 0.25017 & 0.85653 & 0.00474 & 0.99995 & 0.02790 & 0.99822 & 0.00918 & 0.99981 \\
\hline
\end{tabular}
}
\caption{Performance comparison of Vanilla, PINN, Denoiser-EBM, and Denoiser-Fisher models in predicting the Simple Harmonic Oscillator (SHO) solution under increasing levels of Gaussian noise. Metrics reported include RMSE and $R^2 score$.}
\label{tab:model_noise_comparison}
\end{table}
The comparative analysis under varying Gaussian noise levels shows that the Vanilla model exhibits increasing RMSE and decreasing \( R^2 scores\) as noise intensifies. This highlights its its vulnerability to noise which results in poor generalization. The PINN model maintains a constant RMSE and \( R^2 score\) across all noise levels. This is expected, as the PINN does not use noisy observations as input but rather learns the solution using physical variables and the governing equations. Both Denoiser-Fisher and Denoiser-EBM models exhibit stable and reliable performance across all noise levels, with Denoiser-Fisher achieving the best results. While the problem is relatively simple, the consistent outcomes of the denoiser models emphasize the benefit of integrating regularization when learning from noisy data.

\subsection{Burgers' Equation}
The second numerical example is for solving a non-linear 1D time-dependent Burgers' equation~\cite{BURGERS1948171}. Burgers' equation has been used as a benchmark problem for PINN training and evaluation of its performance \cite{RAISSI2019686}. We introduced noise in the solution of the Burgers' equation to recover the analytical solution using denoiser framework varying different noise level.

The differential form of the 1D time-dependent Burgers' equation is given by:

\begin{equation}
    \frac{du}{dt} + u\frac{du}{dx} = v\frac{d^2u}{dx^2}
    \label{eq:burgers' equation}
\end{equation}

Here, \(x\) and \(t\) denote spatial and temporal coordinates, and \(u(x, t)\) is the velocity field dependent on both. The viscosity coefficient \(v\) is set to \(\frac{0.01}{\pi}\). For this simulation, the initial condition is \(u(x, 0) = -\sin(\pi x)\), and Dirichlet boundary conditions are applied: \(u(-1, t) = u(1, t) = 0\). 

\begin{figure}[htbp]
    \centering
    \includegraphics[width=1.0\linewidth]{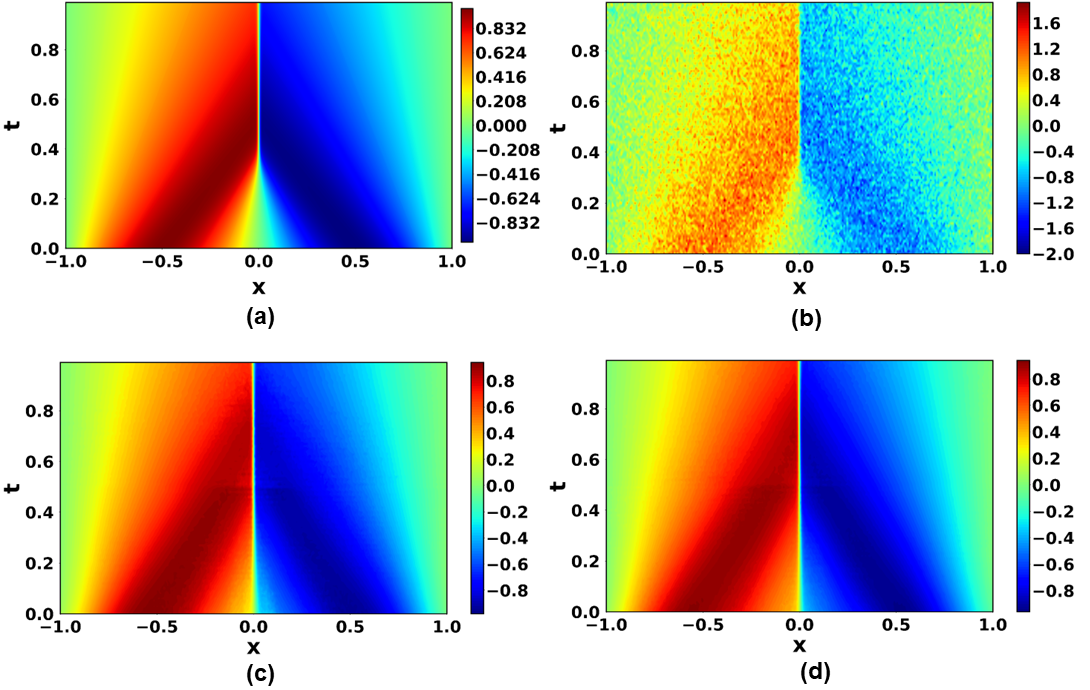}
    \caption{Comparison of the reference and denoised solutions for the Burgers’ equation. (a) Analytical reference solution (ground truth); (b) Ground truth corrupted with 25\% Gaussian noise; (c) Denoised prediction from Denoiser-EBM; (d) Denoised prediction from Denoiser-Fisher.}
    \label{fig:burgers-ref-noisy}
\end{figure}

We used the analytical solution from reference benchmark~\cite{lu2021deepxde}, as shown in Figure~\ref{fig:burgers-ref-noisy} (a) and used it to evaluate the performance of the model. Similar to the SHO problem,  25\% Gaussian noise was added to the reference solution, producing the noisy dataset shown in Figure~\ref{fig:burgers-ref-noisy} (b). Figures~\ref{fig:burgers-ref-noisy} (c) and (d) illustrate the denoised predictions generated by the Denoiser-EBM and Denoiser-Fisher models, respectively. The visual comparison highlights the ability of both denoiser networks to recover the underlying structure of the solution, with Denoiser-Fisher preserving finer details of the reference solution. These results emphasize the importance of incorporating physical constraints or regularization strategies when learning from noisy spatio-temporal data.

\begin{figure}[htbp]
    \centering
    \includegraphics[width=1.0\linewidth]{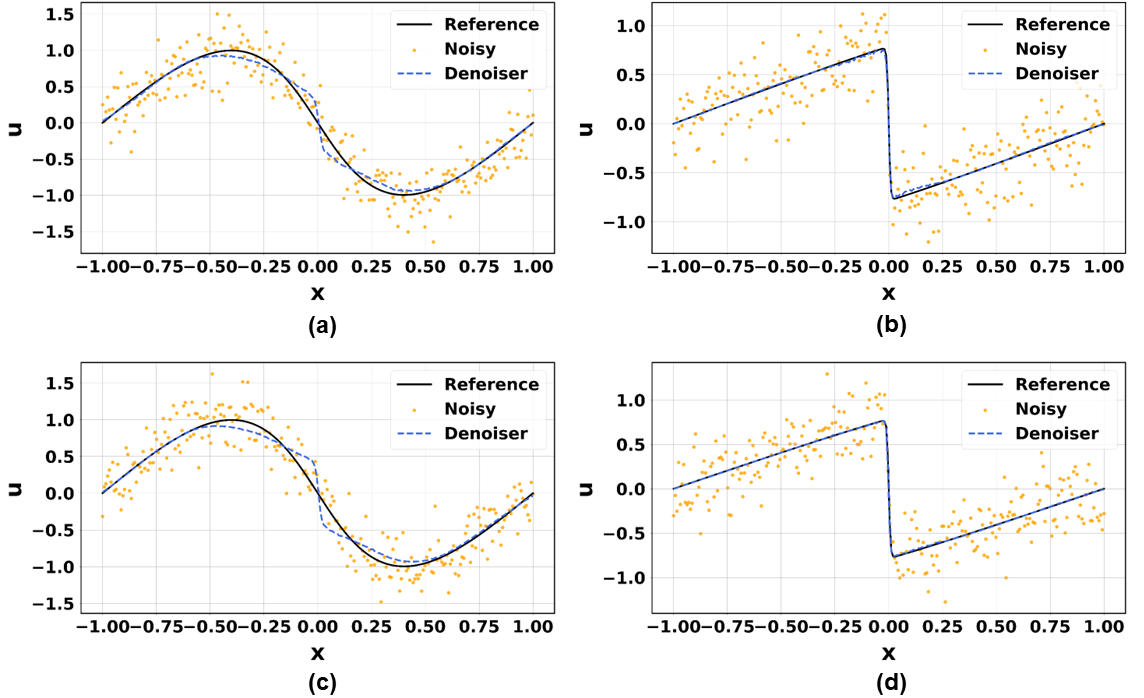}
    \caption{Denoised temporal profiles for the Burgers’ equation at selected time instances. (a), (b) Solutions at \( t = 0.1 \) and \( t = 0.9 \), obtained using the Denoiser-EBM model. (c), (d) Solutions at \( t = 0.1 \) and \( t = 0.9 \), obtained using the Denoiser-Fisher model.}
    \label{fig:burgers-temporal}
\end{figure}

To compare the denoising performance of the proposed models, the predicted profiles of the Burgers’ equation at two representative time instances, \( t = 0.1 \) and \( t = 0.9 \), were examined. Figure~\ref{fig:burgers-temporal} presents the denoised solutions recovered by the Denoiser-EBM and Denoiser-Fisher models, compared against the noisy observations and the ground truth reference solution. Figures 4(a) and 4(b) correspond to the Denoiser-EBM predictions at \( t = 0.1 \) and \( t = 0.9 \), respectively, while Figures~\ref{fig:burgers-temporal} (c) and (d) show the results from the Denoiser-Fisher model at the same time points. Both models demonstrate strong denoising performance over time, with Denoiser-Fisher producing slightly closer fit to the reference curve, particularly near discontinuities and sharp gradients. These temporal profiles validate the efficacy proposed denoisers in restoring physically consistent solutions from corrupted inputs across the spatio-temporal domain.

\begin{table}[htbp]
\centering
\renewcommand{\arraystretch}{1.3}
\resizebox{\textwidth}{!}{%
\begin{tabular}{|c|cc|cc|cc|cc|}
\hline
\textbf{Noise Level} & \multicolumn{2}{c|}{\textbf{Vanilla}} & \multicolumn{2}{c|}{\textbf{PINN}} & \multicolumn{2}{c|}{\textbf{Denoiser-EBM}} & \multicolumn{2}{c|}{\textbf{Denoiser-Fisher}} \\
\cline{2-9}
 & \textbf{RMSE} & \( R^2 \) \textbf{score} & \textbf{RMSE} & \( R^2 \) \textbf{score} & \textbf{RMSE} & \( R^2 \) \textbf{score} & \textbf{RMSE} & \( R^2 \) \textbf{score} \\
\hline
5\%  & 0.05381 & 0.99237 & 0.05887 & 0.99082 & 0.05121 & 0.99305 & 0.05662 & 0.99151 \\
10\% & 0.10105 & 0.97357 & 0.05887 & 0.99082 & 0.05129 & 0.99303 & 0.05663 & 0.99150 \\
15\% & 0.14978 & 0.94366 & 0.05887 & 0.99082 & 0.05123 & 0.99305 & 0.05668 & 0.99149 \\
20\% & 0.19891 & 0.90465 & 0.05887 & 0.99082 & 0.05146 & 0.99298 & 0.05664 & 0.99150 \\
25\% & 0.24820 & 0.85891 & 0.05887 & 0.99082 & 0.05150 & 0.99297 & 0.05673 & 0.99147 \\
\hline
\end{tabular}%
}
\caption{Performance comparison of models on the Burgers’ equation with varying Gaussian noise levels.}
\label{tab:burgers-performance}
\end{table}

Table ~\ref{tab:burgers-performance} summarizes the performance of four models (Vanilla, PINN, Denoiser-EBM, Denoiser-Fisher) on the Burgers’ equation under different levels of Gaussian noise ranging from 5\% to 25\%. The Vanilla model performs consistently poorly, showing relatively high RMSE and low \( R^2 scores\) with increasing noise, indicating its limited ability to generalize under noisy conditions. The PINN model, which does not use noisy data as input but leverages physical laws, consistently achieves the lowest RMSE and highest \( R^2 scores\)  across all noise levels. This highlights the importance of incorporating the noisy data in the training process for better generalization of the model. Denoiser-EBM shows very stable performance across all noise levels, achieving RMSE around 0.051 and \( R^2 scores\) around 0.993, suggesting a reliable denoising capability. Denoiser-Fisher also maintains high \( R^2 scores\) values (above 0.991), but its RMSE is slightly higher than EBM, indicating a small trade-off between smoothness and sharpness.

\subsection{Laplace Equation}
Laplace equation is a canonical second-order partial differential equation (PDE) widely used in mathematical physics to model steady-state phenomena such as heat diffusion. In this study, it serves as a benchmark for evaluating the performance of denoising models in reconstructing smooth, spatially varying fields with localized source disturbances. Its mathematical simplicity and well-known analytical solution properties make it an ideal testbed for our denoising frameworks.

The two-dimensional Laplace equation with a source term is given by ~\cite{ZHANG2022114414}:
\begin{equation}
    \nabla^2 u(x, y) = \frac{\partial^2 u}{\partial x^2} + \frac{\partial^2 u}{\partial y^2} = -b(x, y)
\end{equation}

The source term \( b(x, y) \) is defined as:
\begin{equation}
\begin{aligned}
    b(x, y) &= \left(20 - 400 (x - 5)^2\right) e^{-10 (x - 5)^2} 
    \left(e^{-10 (y - 5)^2} - e^{-250} \right) \\
    &\quad + \left(e^{-10 (x - 5)^2} - e^{-250} \right) 
    \left(20 - 400 (y - 5)^2\right) e^{-10 (y - 5)^2}
\end{aligned}
\end{equation}
Here, \( u(x, y) \) represents the unknown scalar field (e.g., temperature or potential), and \( b(x, y) \) introduces a spatially concentrated source near the center of the domain. The PDE is solved over a bounded domain \( \Omega \), with homogeneous Dirichlet boundary conditions: \( u(x, y) = 0 \) for all \( (x, y) \in \partial \Omega \).

An analytical reference solution adapted from ~\cite{ZHANG2022114414} is used for benchmarking:

\begin{equation}
    u_{\text{true}}(x, y) = \left(e^{-10(x-5)^2} - e^{-250}\right) \left(e^{-10(y-5)^2} - e^{-250}\right)
\end{equation}

\begin{figure}[htbp]
    \centering
    \includegraphics[width=1.0\linewidth]{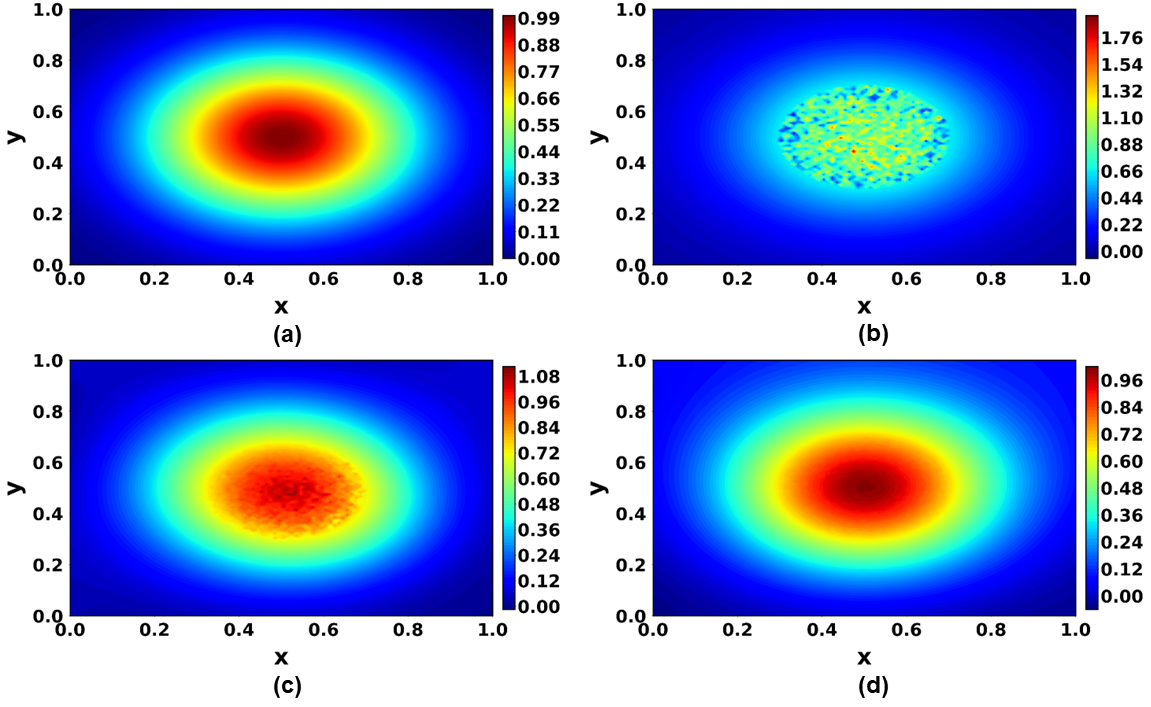}
    \caption{Comparison of analytical, noisy, and denoised solutions for the 2D Laplace equation with a centrally concentrated source. (a) Ground truth; (b) Noisy input (25\% Gaussian noise); (c) denoised solution from Denoiser-EBM; (d) denoised solution from Denoiser-Fisher.}
    \label{fig:laplace-visual}
\end{figure}

To simulate measurement imperfections, 25\% Gaussian noise is added to the analytical solution (Figure~\ref{fig:laplace-visual}) around the concentrated source. All models are trained on this noisy dataset. Additional tests are conducted at 5\%, 10\%, 15\%, and 20\% noise levels to evaluate generalization.

\begin{figure}[ht]
    \centering
    \includegraphics[width=1.0\linewidth]{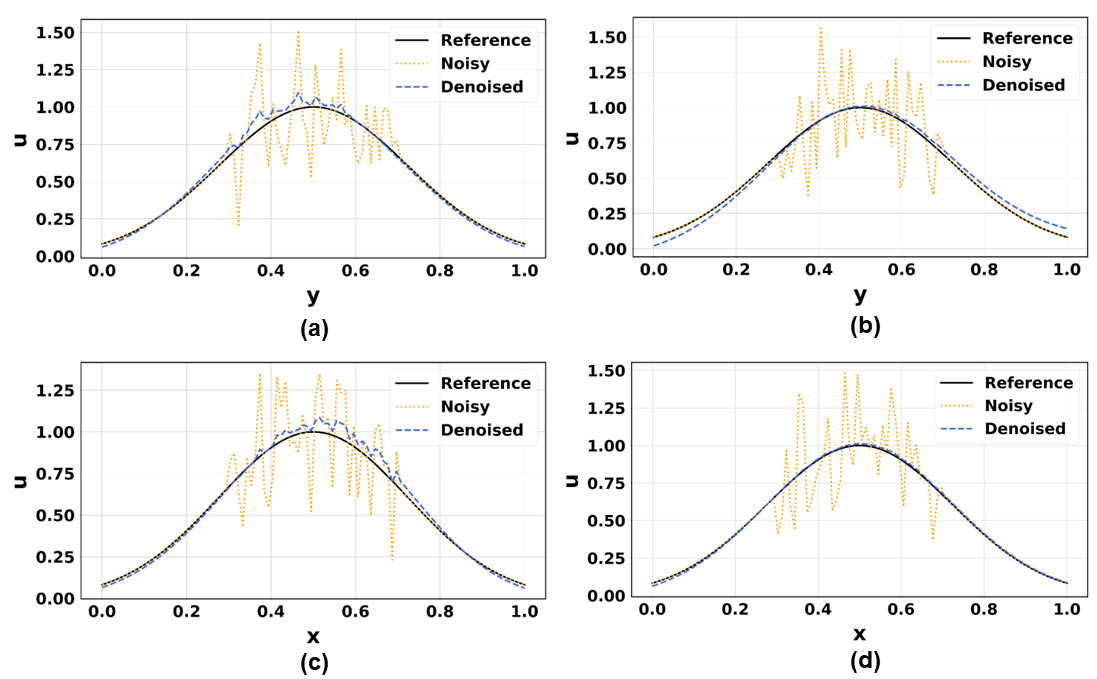}
    \caption{Cross-sectional slices of the denoised Laplace solution at \( x = 0.5 \)  for (a) Denoiser-EBM and (b) Denoiser-Fisher model; solution at \( y = 0.5 \)  (c) Denoiser-EBM and (d) Denoiser-Fisher.}
    \label{fig:laplace_slices}
\end{figure}

Spatial cross-sectional profiles in Figure~\ref{fig:laplace_slices} compare the learned outputs with the true solution along horizontal and vertical midlines. Both Denoiser-EBM and Denoiser-Fisher recover fine-scale features, with Denoiser-Fisher producing smoother results in high-noise regions.

\begin{table}[htbp]
\centering
\caption{Comparison of RMSE and $R^2$ scores for different models across varying levels of Gaussian noise.}
\resizebox{\textwidth}{!}{%
\begin{tabular}{|c|cc|cc|cc|cc|}
\hline
\textbf{Noise Level} 
& \multicolumn{2}{c|}{\textbf{Vanilla}} 
& \multicolumn{2}{c|}{\textbf{PINN}} 
& \multicolumn{2}{c|}{\textbf{Denoiser-EBM}} 
& \multicolumn{2}{c|}{\textbf{Denoiser-Fisher}} \\
\cline{2-9}
 & RMSE & $R^2$ score 
 & RMSE & $R^2$ score 
 & RMSE & $R^2$ score 
 & RMSE & $R^2$ score \\
\hline
5\%  & 0.051295 & 0.962219 & 0.046295 & 0.968228 & 0.023080 & 0.979883 & 0.036838 & 0.992104 \\
10\% & 0.100721 & 0.869823 & 0.046295 & 0.968228 & 0.023420 & 0.979883 & 0.036838 & 0.991869 \\
15\% & 0.151110 & 0.742724 & 0.046295 & 0.968228 & 0.023493 & 0.979882 & 0.036842 & 0.991818 \\
20\% & 0.199379 & 0.626618 & 0.046295 & 0.968228 & 0.024240 & 0.979882 & 0.036849 & 0.991290 \\
25\% & 0.249846 & 0.524780 & 0.046295 & 0.968228 & 0.024322 & 0.979867 & 0.036853 & 0.991231 \\
\hline
\end{tabular}
}
\label{tab:laplace-metrics}
\end{table}

As shown in Table~\ref{tab:laplace-metrics}, the Vanilla model consistently performs worst under noise, with high RMSE and low $R^2 scores$  (e.g., RMSE = 0.249846 at 25\%). The PINN improves upon this by leveraging physical constraints. In contrast, Denoiser-EBM and Denoiser-Fisher deliver robust performance across all noise levels. Denoiser-EBM achieves the lowest RMSE and Denoiser-Fisher achieves highest $R^2 scores$  overall, showing strong generalization with stable metrics.

\subsection{Discussion on numerical examples}
These results highlight the importance of physics-based and statistical models in learning from noisy data. While the PINN model improves robustness by enforcing physical laws, its performance still degrades with increasing noise. In contrast, the Denoiser-EBM introduces an energy-based prior that guides the model toward physically consistent solutions, achieving the lowest RMSE and highest \( R^2 scores\)  across all noise levels. The Denoiser-Fisher model, using Fisher information-based regularization, further enhances prediction stability and smoothness, especially under spatially varying noise. Both denoiser models demonstrate superior noise resilience compared to the Vanilla network, which lacks any form of regularization. These findings support the use of physics-guided denoiser frameworks that integrate physics and statistics for solving noisy problems in engineering applications.

\section{Physics-Informed Neural Network for LPBF Simulation}
\subsection{Problem Definition}
LPBF is a thermally-driven process where various sensors—such as infrared (IR) cameras, photodiodes, and other thermal sensors—are used to monitor temperature fields during printing. This thermal data is critical for understanding melt pool dynamics, microstructure evolution, and defect formation in the printed part. While physics-based modeling and simulation can provide valuable insights into the process and aid in designing defect-minimized strategies, they are often computationally expensive and impractical for real-time prediction, especially for guiding the printing of the next layer. Moreover, such simulations require extensive calibration and are sensitive to discrepancies caused by noisy experimental data, making validation and deployment in real-world scenarios challenging.
Given the rapid and complex interactions between the laser and the powder bed, physics-based approaches combined with statistical models offer a promising strategy to interpret and learn the underlying thermal behavior. In the previous section, we demonstrated that such models can effectively learn from biased and noisy synthetic data. In this section, we extend this framework to real-world noisy dataset from LPBF process.

\begin{figure}[htbp]
    \centering
    \includegraphics[width=0.9\linewidth]{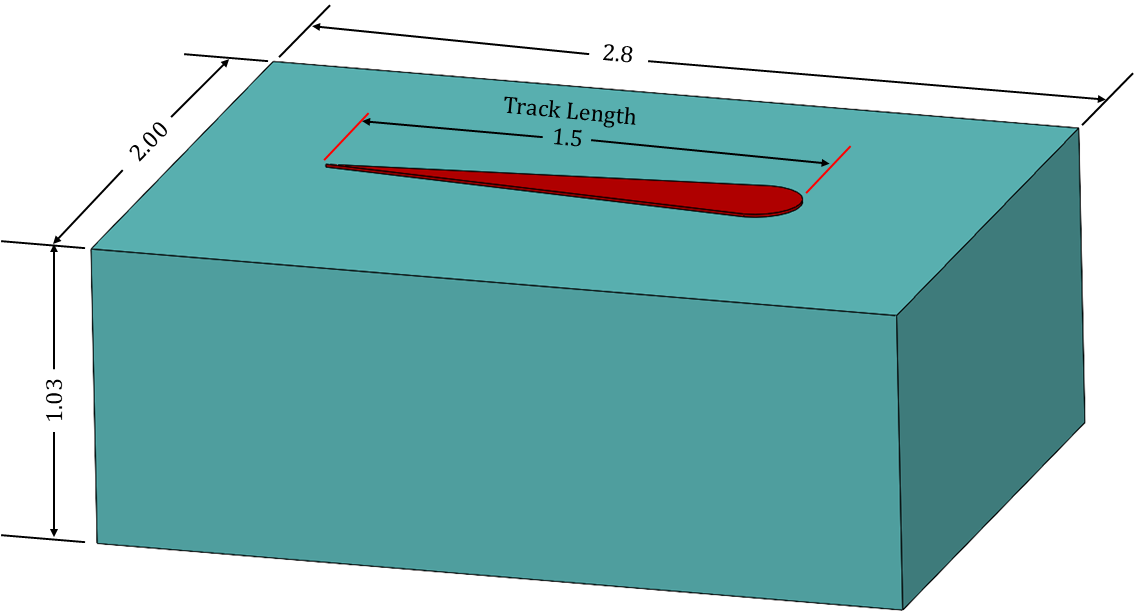}
    \caption{Domain geometry used for LPBF single-track simulation, adapted from \cite{HOSSEINI2023116019}.}
    \label{fig:domain}
\end{figure}

To train our physics-guided denoiser network, we first need to establish a physics-based model. Here, we trained a PINN -AM model for LPBF simulation. Later, we incorporate this pretrained PINN-AM model to the denoiser network . 
The PINN-AM model is trained for a 3D simulation of  single track scan of the LPBF process  varying a wide range of laser power and scan speed. The physical domain has a dimension of  \(2.8 \times 2.0 \times 1.03\, \text{mm}^3\), representing the build volume in a single-track LPBF simulation with a layer height of 0.03 mm. A schematic of the simulation domain is shown in Figure \ref{fig:domain}
. 


The PINN-AM is trained to approximate the temperature field that satisfies the transient heat conduction equation with a volumetric heat source model:
\begin{equation}
\rho_0 \tilde{C}_p \dot{T} = \nabla \cdot \left( \tilde{k} \nabla T \right) + q_{\text{vol}}
\end{equation}

Here, \( \rho_0 \), \( \tilde{C}_p \), and \( \tilde{k} \) represent the reference density, apparent specific heat capacity, and effective thermal conductivity, respectively. The term \( q_{\text{vol}} \) accounts for the volumetric heat input from the laser. The apparent heat capacity \( \tilde{C}_p \) differs from the material's intrinsic heat capacity \( C_p \) in regions where phase transformations occur, incorporating the associated latent heat.

The volumetric heat source term is modeled using a semi-ellipsoidal Gaussian volumetric heat source, commonly employed in laser-based additive manufacturing simulations:

\begin{equation}
q_{\text{vol}} = \alpha \frac{6 \sqrt{3} P}{\pi \sqrt{\pi} r^3} 
\exp\left( -3 \frac{(x + vt)^2 + y^2}{r^2} \right) 
\exp\left( -3 \frac{z^2}{c^2} \right)
\end{equation}

where \( P \) is the laser power, \( \alpha \) is the laser absorption coefficient, \( r \) is the laser beam radius, \( c \) is the laser penetration depth, and \( v \) is the laser scan speed. The initial temperature of the domain is fixed at \(25^\circ\)C, and heat losses due to convection and radiation from the top surface are neglected to simplify the modeling. 

\subsection{PINN-AM model training}
To solve the temperature evolution under varying process and material conditions, we trained a PINN-AM model following previous work by ~\cite{HOSSEINI2023116019}. 

The PINN architecture consists of a fully connected feed-forward neural network with six hidden layers, each containing 24 neurons and using $sine$ activation functions. This configuration is consistent with the architecture employed by ~\cite{HOSSEINI2023116019}. The network takes eight inputs: three spatial coordinates \((x, y, z)\), time \(t\), laser power multiplied by absorption coefficient \((P \times \alpha)\), laser scan speed \(v\), thermal conductivity \(k\), and specific heat capacity multiplied by reference density \((\rho_0 \tilde{C}_p)\). The output is the predicted temperature \(T(x, y, z, t, k, \rho_0 \tilde{C}_p, P \alpha, v)  \).

\begin{figure}[ht]
\centering
\scalebox{0.82}{
\begin{tikzpicture}[x=1.4cm, y=0.45cm]

\tikzstyle{neuron}=[circle, draw=black!70, fill=white, minimum size=5mm, inner sep=0pt]
\tikzstyle{input neuron}=[circle, draw=black, thick, fill=gray!10, minimum size=10mm]
\tikzstyle{output neuron}=[circle, draw=black, thick, fill=white, minimum size=10mm]
\tikzstyle{connection}=[-stealth, thin, draw=black!20]
\tikzstyle{layer label} = [font=\footnotesize, above]

\foreach \name/\y/\label in {
  x/24/$x$,
  y/21/$y$,
  z/18/$z$,
  t/15/$t$,
  pa/12/$P\alpha$,
  v/9/$v$,
  k/6/$k$,
  cp/3/{$\rho C_p$}
}{
  \node[input neuron] (\name) at (0,\y) {\label};
}
\node[align=center] at (-0.9, 13) {\textbf{Inputs}};

\def\nlayers{6}
\def\nneurons{24}

\foreach \l in {1,...,\nlayers}{
  \foreach \n in {1,...,\nneurons}{
    \pgfmathsetmacro\y{(\n-1)*28/(\nneurons-1)} 
    \node[neuron] (H\l_\n) at (\l,\y) {};
  }
  \node[layer label] at (\l,29) {$H_{\l}$};
}

\node[output neuron] (out) at ({\nlayers+1},13.5) {$T$};
\node at ({\nlayers+2.0}, 13.5) {\textbf{Output}};

\foreach \src in {x,y,z,t,pa,v,k,cp}{
  \foreach \n in {1,3,...,23}{  
    \draw[connection] (\src.east) -- (H1_\n.west);
  }
}

\foreach \l [evaluate=\l as \next using int(\l+1)] in {1,...,\numexpr\nlayers-1}{
  \foreach \i in {1,...,\nneurons}{
    \foreach \j in {1,3,...,23}{  
      \draw[connection] (H\l_\i.east) -- (H\next_\j.west);
    }
  }
}

\foreach \n in {1,3,...,23}{
  \draw[connection] (H\nlayers_\n.east) -- (out.west);
}


\def\ycenter{12.0}
\def\yupper{18.8}
\def\ylower{6.2}

\def\boxleft{8.5}
\def\boxright{13.0}
\def\boxheight{0.6}  

\draw[thick, rounded corners] (\boxleft,\ycenter-\boxheight) rectangle (\boxright,\ycenter+\boxheight);
\node[align=left] at (10.75,\ycenter) {\normalsize $\rho_0 \tilde{C}_p \dot{T} = \nabla \cdot (\tilde{k} \nabla T)  + q_{\text{vol}}$};
\draw[->, thick] (out) -- (\boxleft,\ycenter);

\draw[thick, rounded corners] (\boxleft,\yupper-\boxheight) rectangle (\boxright,\yupper+\boxheight);
\node[align=left] at (10.75,\yupper) {\normalsize $T(x,y,z,0) = T_0$ \quad \textit{(IC)}};
\draw[->, thick] (out) -- (\boxleft,\yupper);

\draw[thick, rounded corners] (\boxleft,\ylower-\boxheight) rectangle (\boxright,\ylower+\boxheight);
\node[align=left] at (10.75,\ylower) {\normalsize $T|_{\partial \Omega} = T_b$ \quad \textit{(BC)}};
\draw[->, thick] (out) -- (\boxleft,\ylower);

\end{tikzpicture}
}
\caption{PINN-AM architecture for LPBF thermal analysis with 8 inputs, 6 hidden layers, and physics-based training. Annotations show the governing heat equation, initial condition (IC), and boundary condition (BC).}
\end{figure}
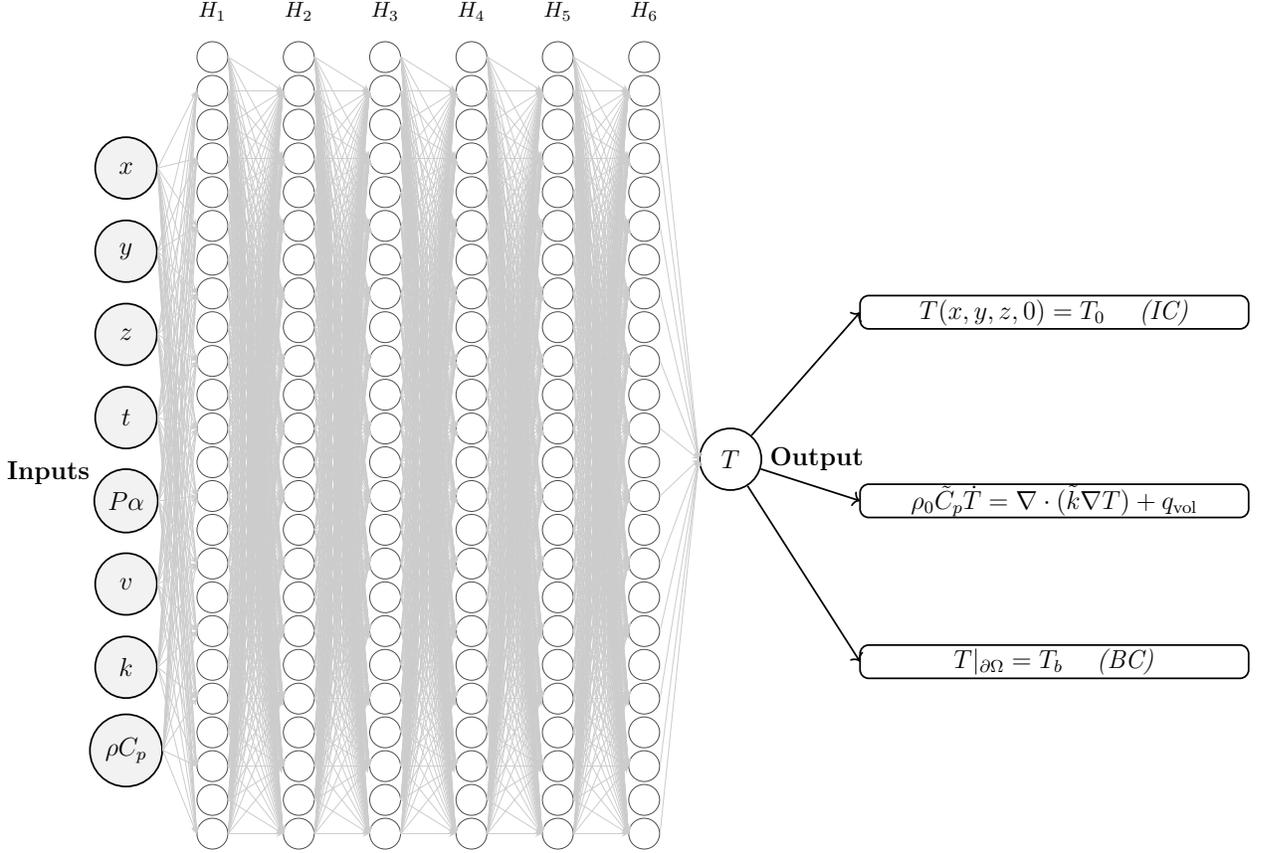

The training process involves minimizing a physics-based loss function is formulated based on the governing equations above and optimized using the L-BFGS optimizer over \(2^{19}\) collocation points. These points are generated using a hybrid sampling strategy that combines spherical random sampling with a Cartesian Sobol sequence to ensure diverse coverage of the spatiotemporal domain. Additionally, boundary conditions are enforced using \(2^{18}\) points sampled from the domain boundaries. To ensure numerical stability and convergence, the neural network is configured with 6 hidden layers, each containing 24 neurons, and uses the $sine$ activation function to better capture the periodic and nonlinear features of the temperature field. The model is trained for a maximum of 50,000 L-BFGS iterations, which allows for efficient quasi-Newton optimization of the full-batch loss. The density is fixed at the reference value of \(\rho_0 = 8352 \, \text{kg/m}^3\), representative of Hastelloy X at room temperature, to ensure mass conservation within the simulated build volume.

\subsection{Validation of PINN-AM}
\begin{figure}[ht]
    \centering
    \includegraphics[width=0.8\linewidth]{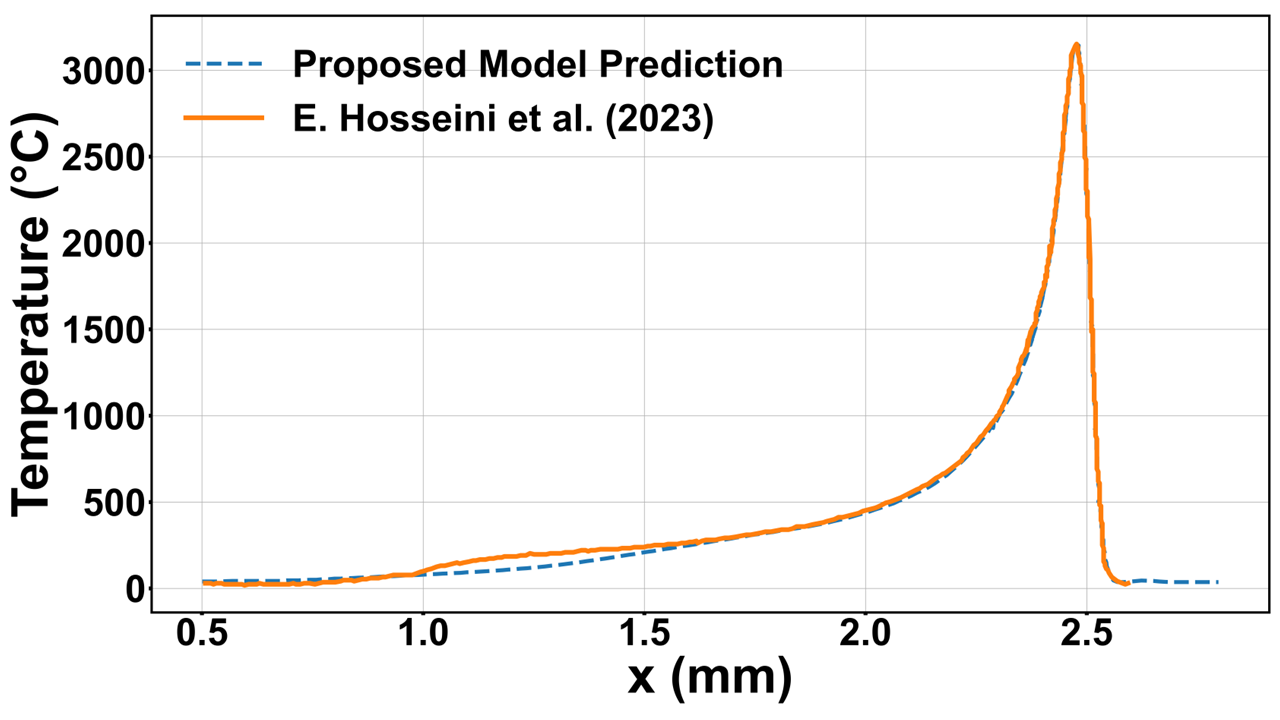}
    \caption{Validation of the proposed model against benchmark data from ~\cite{HOSSEINI2023116019}. The predicted temperature profile (dashed blue) is compared with the reference profile (solid orange). Evaluation metrics indicate strong agreement between the two.}
    \label{fig:validation}
\end{figure}

To assess the accuracy of the proposed model, we validate its temperature predictions against benchmark results reported by ~\cite{HOSSEINI2023116019}. Figure~\ref{fig:validation} presents a comparison between the predicted temperature profile generated by our model and the reference solution under identical process parameters, including laser power, scan speed, thermal conductivity, and volumetric heat capacity.

The model successfully captures both the overall thermal trend and the sharp peak near the melt pool center. This agreement is quantitatively supported by a MAE of 45.26, a RMSE of 63.53, and a $R^2$ score of 0.9957, indicating high accuracy and consistency with the benchmark profile.

This pretrained PINN-AM model is used with the denoiser network for two test cases: i) synthetic noise on single track temperature measurement  where ground truth is known as the validated PINN-AM solution, ii) photodiode-based real LPBF noisy measurement (where the ground truth and noise level is unknown)
Before we discuss the results for the two cases, first we describe the experimental data collection for the second test case. 

\section{Experimental noisy data collection}
\begin{figure}[ht]
    \centering
    \includegraphics[width=1.0\linewidth]{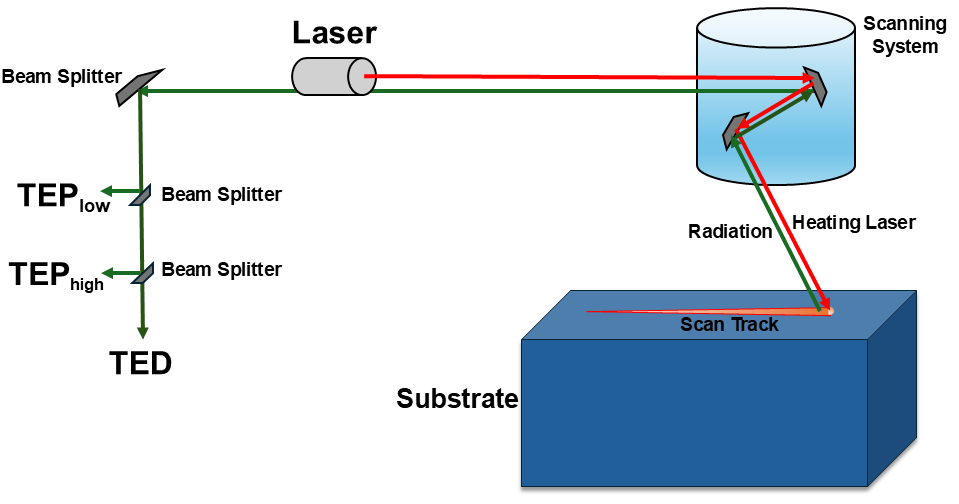}
    \caption{Schematic of the experimental setup for in-situ melt pool monitoring system using photodiodes. Radiation from the melt pool of scan track are split into three channels (TEP High, TEP Low, and TED) for spectral analysis and temperature estimation.}
    \label{fig:exp-setup}
\end{figure}

A schematic of the experimental setup used for in-situ melt pool monitoring is shown in Figure~\ref{fig:exp-setup}. The system is based on the PrintRite3D\textsuperscript{®} platform (Sigma Additive Solutions, USA), which captures high-speed spectral emissions from the melt pool during laser powder bed fusion (LPBF). Emissions are collected using three photodiodes that separately detect high-wavelength, low-wavelength, and broadband signals. These signals are processed to compute two key features: Thermal Emission Planck (TEP\textsuperscript{TM}), and Thermal Energy Density (TED\textsuperscript{TM}). This setup enables real-time acquisition of temperature and energy density data at sampling rates up to 200 kHz, providing high-resolution monitoring of the melt pool dynamics throughout the build process~\cite{porter2022generation}. The experimental dataset used to train the denoiser model comprises four single-track scans. Each track corresponds to a distinct combination of laser power and scan speed, ranging from 240–360~W and 1464–1607~mm/s, respectively. The material used in all builds was AlSi10Mg, with thermal conductivity in the range of 0.207–0.209~W/(m·K)/1000 and volumetric heat capacity between 0.005054–0.005113~kg/mm\textsuperscript{3}·J/(kg·K).

\begin{figure}[ht]
    \centering
    \includegraphics[width=1.0\linewidth]{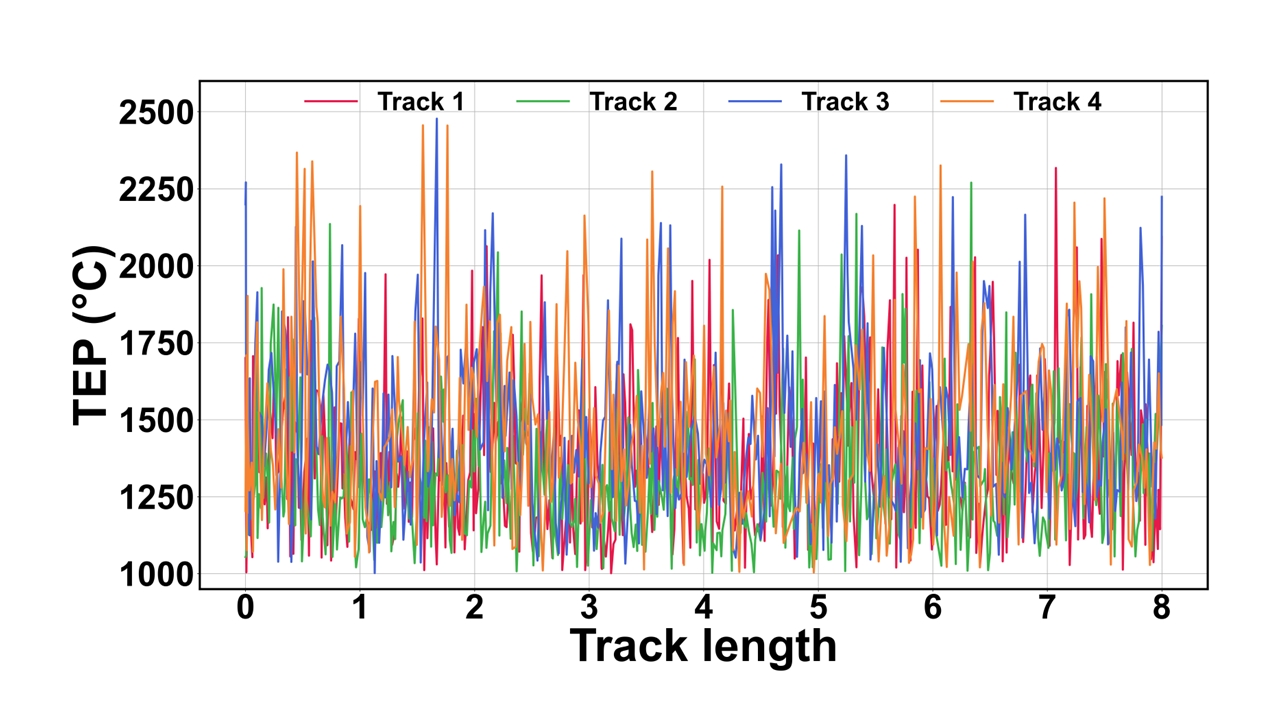}
    \caption{Noisy TEP data across four single laser tracks collected using the photodiodes. These noisy data are used to training the denoiser models for unknown noise level.}
    \label{fig:fig10}
\end{figure}

The collected TEP values were highly noisy, with raw values ranging from as low as –7900 to over 100{,}000 in some cases due to sensor saturation or optical artifacts. To ensure meaningful input to the denoiser model, only TEP values within the physically realistic range of 1000 to 2500 were retained after filtering. For instance, from track 2, 511 valid points remained after removing 21 outliers. Figure~\ref{fig:fig10} illustrates the spatial distribution of the filtered TEP data across the four tracks, color-coded by temperature. These measurements, despite noise, retain critical thermal gradients essential for training the denoiser to robustly infer true temperature distributions under varied process conditions.

\section{Application of denoiser network for LPBF sensing data}

\subsection{Case 1: Synthetic noise with known ground truth and noise level}
To demonstrate the denoising capability of the denoiser network for AM case, we first perform a controlled numerical experiment using the trained PINN-AM model. A synthetic noisy temperature profile is generated by injecting 25\% Gaussian noise into a simulated single-track output from the validated PINN-AM model. The track spans 1.5 mm along the build direction, following the setup of benchmark\#1 from ~\cite{HOSSEINI2023116019}. We also evaluate the denoising performance across multiple noise levels by measuring RMSE and Signal-to-Noise Ratio (SNR) in decibels (dB). The results are summarized in Table~\ref{tab:lpbf_denoise_results}.

\begin{figure}[ht]
    \centering
    \includegraphics[width=1.0\linewidth]{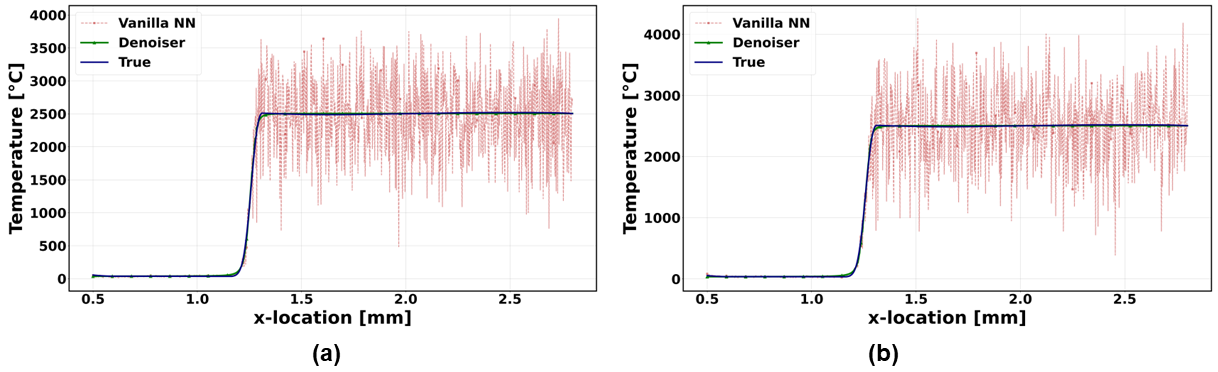}
    \caption{Comparison of denoised temperature profiles against noisy and reference solutions. (a) Denoiser-EBM model and (b) Denoiser-Fisher model results. Both models successfully suppress high-frequency noise while recovering the thermal peak correctly.}
    \label{fig:controlled-denoise}
\end{figure}

\begin{table}[htbp]
\centering
\renewcommand{\arraystretch}{1.3}
\resizebox{\textwidth}{!}{%
\begin{tabular}{|c|ccc|ccc|ccc|}
\hline
\textbf{Noise Level} 
& \multicolumn{3}{c|}{\textbf{Vanilla}} 
& \multicolumn{3}{c|}{\textbf{Denoiser-EBM}} 
& \multicolumn{3}{c|}{\textbf{Denoiser-Fisher}} \\
\cline{2-10}
 & \textbf{RMSE} & \textbf{MAE} & \textbf{SNR (dB)} 
 & \textbf{RMSE} & \textbf{MAE} & \textbf{SNR (dB)} 
 & \textbf{RMSE} & \textbf{MAE} & \textbf{SNR (dB)} \\
\hline
5\%  & 129.551206 & 87.864371  & 12.9625  & 12.975421 & 8.929096  & 20.015565  & 12.960658 & 8.654303  & 19.702969 \\
10\% & 256.634587 & 171.571591 &  9.8540  & 13.026868 & 8.950125  & 17.00533   & 13.073619 & 8.675269  & 18.181588 \\
15\% & 382.780412 & 255.168167 &  8.3076  & 13.191471 & 8.980758  & 15.244492  & 13.462113 & 8.796687  & 17.998887 \\
20\% & 507.230774 & 338.070237 &  6.8717  & 13.680696 & 9.135557  & 13.995185  & 14.16731  & 8.839815  & 16.219233 \\
25\% & 629.507785 & 420.036296 &  6.1316  & 14.414985 & 9.387870  & 13.026173  & 14.169578 & 8.907973  & 14.086949 \\
\hline
\end{tabular}
}
\caption{Comparison of RMSE, MAE, and SNR (in dB) across different Gaussian noise levels for Vanilla, Denoiser-EBM, and Denoiser-Fisher models applied to the LPBF simulation. The results demonstrate the effectiveness of the hybrid denoising models in maintaining low error and higher signal quality under increasing noise.}
\label{tab:lpbf_denoise_results}
\end{table}

Table~\ref{tab:lpbf_denoise_results} presents a comparative evaluation of the Vanilla NN, Denoiser-EBM, and Denoiser-Fisher models under progressively increasing levels of Gaussian noise in the LPBF simulation. The Vanilla model exhibits substantial degradation, with RMSE rising sharply from 103.37 to 498.30 and SNR dropping from 12.96 dB to 6.13 dB. In contrast, both denoiser-based models maintain significantly lower RMSE values and higher SNRs across all noise levels, highlighting their robustness. The Denoiser-EBM shows consistent error levels around 15 RMSE but a gradual decline in SNR, indicating noise sensitivity under large distortion than the denoised signals. Meanwhile, the Denoiser-Fisher model consistently achieves the lowest RMSE and highest SNR for most of the noise levels, demonstrating superior signal and noise suppression. These findings validate the effectiveness and generalization capacity of the proposed hybrid denoising models.

\subsection{Case 2: Experimental LPBF noisy data with unknown ground truth and noise level}
To evaluate the performance of the proposed physics-based denoising framework, we applied it to real single track experimental TEP data. We used multiple single track data with unknown noise level to train the model and predict on a completely new experimental noisy measurement ~\cite{porter2022generation}. The goal was to denoise the noisy thermal measurements and reconstruct smooth, physically consistent temperature profiles across the scan length. The denoiser model was designed as a fully connected feedforward neural network with before mentioned details. Each layer used the \texttt{Tanh} activation function, along with a dropout probability of 0.15 to mitigate overfitting with the final layer outputing a single scalar temperature prediction. The total loss function comprised three components: (1) the data loss, computed as the mean squared error between the denoised prediction and the physics model output; (2) the physics loss ; and (3) an additional regularization loss from either an Energy-Based Model (EBM) or a Fisher Score model. The EBM and Fisher Score networks were both constructed as four-layer neural networks with 24 hidden units per layer and Tanh activations, designed to model uncertainty through log-partition-based energy and Fisher information, respectively.

We employed 4-fold cross-validation (CV) to assess the generalization. In each fold, the model was trained for 20{,}000 epochs using the ADAM optimizer with a cosine annealing learning rate scheduler. The best-performing fold (Fold 3) achieved a CV RMSE of 26.49, while the overall average CV RMSE was 33.46. After cross-validation, we conducted full training on the combined dataset to produce the final denoiser model.

\begin{figure}[ht]
    \centering
    \includegraphics[width=1.0\linewidth]{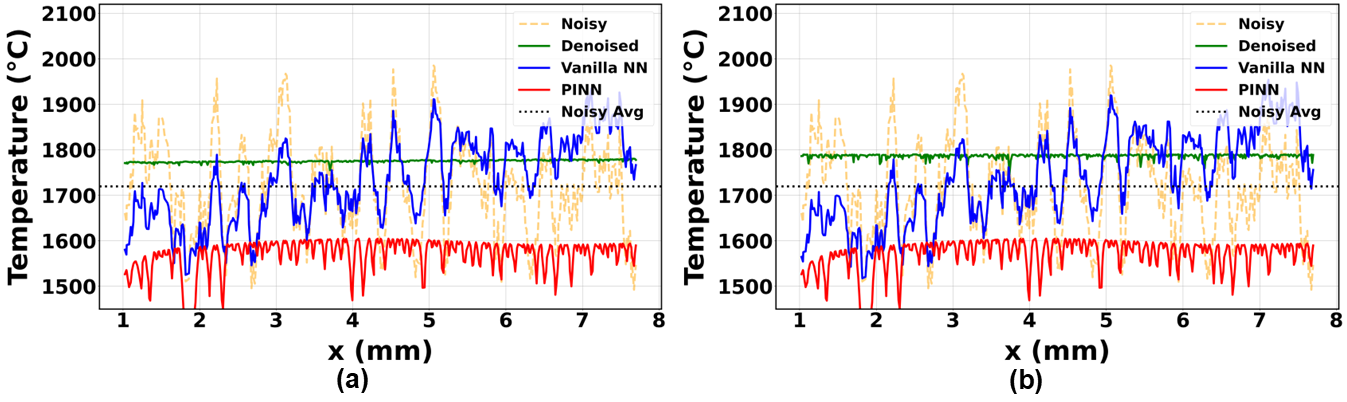}
    \caption{Comparison of denoiser models trained with different uncertainty regularization strategies on noisy TEP data: (a) Denoiser with EBM loss, (b) Denoiser with Fisher Score loss. Results are compared against noisy input data and a vanilla neural network baseline.}
    \label{fig:denoiser-comparison}
\end{figure}

Figure~\ref{fig:denoiser-comparison} presents a visual comparison of model predictions on a representative track. The left subplot (a) shows results from the EBM-regularized denoiser, while the right subplot (b) shows those from the Fisher Score-regularized denoiser. In both cases, the denoised output (green curve) closely follows a physically plausible trend, effectively suppressing noise compared to the raw input (dashed gray). In contrast, the Vanilla neural network model was trained solely on the four experimental noisy datasets using MSE loss between its output and the noisy data. When evaluated on the denoised outputs from both models, it produces overfitted, high-variance predictions (blue curve). While the Denoiser-EBM model slightly underpredicts peak temperature values, it maintains smoothness and global consistency. The Denoiser-Fisher model captures the shape of the thermal profile more accurately, showing better agreement with expected temperature behavior. These results validate the effectiveness of the physics- and uncertainty-informed denoising framework for enhancing thermal field prediction in metal additive manufacturing.

Table~\ref{tab:metrics} presents the performance comparison among the three models using RMSE, SNR, and MAE. Both denoising models, Denoiser-EBM and Denoiser-Fisher, consistently outperform the Vanilla Neural Network. The denoisers achieve lower RMSE and MAE values, indicating better accuracy and reduced error in temperature prediction. While the improvement in SNR is relatively modest due to the absence of ground truth in experimental data, both denoisers maintain slightly higher SNR values, reflecting improved robustness against noise. Among them, Denoiser-Fisher shows the best performance, particularly excelling in producing smoother and more reliable outputs.

\begin{table}[htbp]
\centering
\renewcommand{\arraystretch}{1.3}
\resizebox{0.5\textwidth}{!}{%
\begin{tabular}{|c|c|c|c|}
\hline
Metrics & Vanilla & Denoiser-EBM & Denoiser-Fisher \\
\hline
RMSE     & 532.54 & 491.51 & 462.25 \\
MAE      & 321.03 & 317.69 & 314.00 \\
SNR (dB) & 12.56  & 13.10  & 13.09 \\
\hline
\end{tabular}
}
\caption{Comparison of model performance on experimental temperature data using RMSE, MAE, and SNR.}
\label{tab:metrics}
\end{table}

\section{Conclusion}
In this study, we developed a physics-guided denoising framework using PINN as a physics engine and EBM and Fisher for statistical noise estimation. We impleneted the physics-guided denoiser framework for three numerical examples of SHO, Burgers' equation, and Laplace equation. The trained denoiser networks have maintained a consistent performance up to 25\% noise level both the denoiser-EBM and denoiser-Fisher. We used a baseline of Vanilla neural network and standard PINN to compare both the denoiser-EBM and denoiser-Fisher model performance. The denoiser-Fisher generally performed better compared to denoiser-EBM  and provide a smooth solution for a localized noise case of Laplace equation. Further, we trained a PINN-AM model for a physics-based simulation of the LPBF process and validated it against a finite element model for temperature prediction. We later used this pretrained PINN-AM model in the denoiser network as the physics engine for two scenarios: i)  where the ground truth temperature and noise level is known and ii) real experimental case where the ground truth and noise level is unknown. For both cases, denoiser-EBM and denoiser-Fisher effectively reduced the noise compared to Vanilla or standard PINN which is  established through metrics such as RMSE and SNR. 


While the proposed physics-guided denoiser is robust, the model is not tested under wide experimental range due to the scarcity of reliable, high-resolution experimental datasets, which could affect the generalization of the denoiser models for LPBF. To address this, we plan to extend our work to multi-track and multi-layer scenarios, which better reflect real-world additive manufacturing processes. Future directions will also include integrating multiple real-time sensor feedback, and incorporating domain-specific constraints to further improve prediction accuracy and reliability. With continuous training with new sensor data, the physics-guided denoiser could become a key tool for cleaning real-time AM data, enabling more accurate process predictions and improving defect control through predictive process adjustments.

\vspace{0.5 cm}

\textbf{Acknowledgments}\\
S.M. and P.H. gratefully acknowledge the start-up fund from Washington State University to support this research. The authors thank Prof. Jian Cao (Northwestern University) for providing the single track MPM experimental data. 

\vspace{0.5 cm}

\textbf{Data availability}\\
Data generated and used for this research will be made available upon reasonable request. 

\vspace{0.5 cm}

\textbf{Declaration of competing interest}\\
The authors declare that they have no known competing financial interests or personal relationships that could have appeared to influence the work reported in this paper.


 \bibliographystyle{elsarticle-num}
 \bibliography{manuscript.bib}





\end{document}